\documentclass{PoS}

\def\MeV{{\rm Me\!V}}

\title{Recent progress in finite temperature lattice QCD}

\ShortTitle{Recent progress in finite temperature lattice QCD}

\author{\speaker{Urs M. Heller}\\
        American Physical Society, One Research Road, Box 9000,
        Ridge, NY 11961, USA\\
        E-mail: \email{heller@aps.org}}

\abstract{
I review recent progress in the determination of the QCD phase diagram
at finite temperature, in investigations of the nature of the transition
or crossover from the hadronic phase to the quark-gluon plasma phase
and in the determination of the equation of state. This talk will
focus on results at zero chemical potential.
}

\FullConference{XXIVth International Symposium on Lattice Field Theory\\
                July 23-28, 2006\\
                Tucson, Arizona, USA}

\begin{document}

\section{Introduction}

\begin{figure}[h]
\centering
\includegraphics[width=8cm]{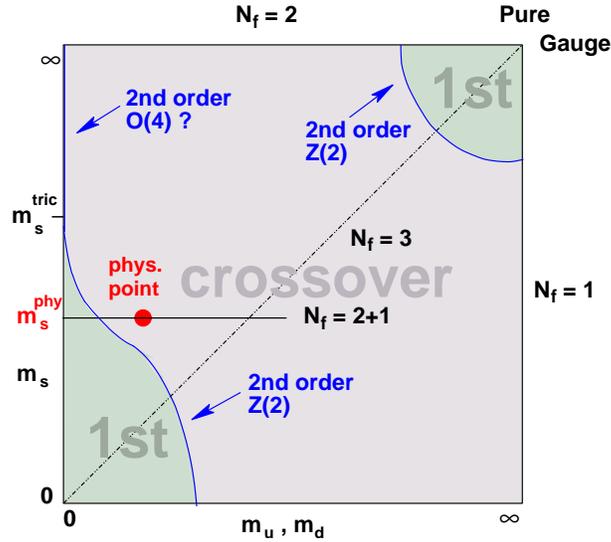}
\caption{\label{ud_s_plane} Sketch of the QCD phase diagram in the
$m_{ud} - m_s$ plane.}
\end{figure}

In the past few years our understanding of QCD at finite temperature has
advanced considerably, due to improvements in the lattice discretizations
(use of improved actions), improvements in the simulation algorithms,
and increases in computational resources. A sketch of our present
understanding of the QCD phase diagram in the $m_{ud} - m_s$ plane is
shown in Fig.~\ref{ud_s_plane}.

In this talk I review recent progress in the determination of the QCD phase
diagram at finite temperature and zero baryon density, in investigations
of the nature of the transition or crossover from the hadronic phase at
low temperatures to the quark-gluon plasma (QGP) phase at high temperatures
and in the determination of the equation of state (EOS). For earlier recent
reviews see Refs.~\cite{Reviews}, while a summary of other properties
of the high-temperature phase is given in the talk by
T.~Hatsuda~\cite{Hatsuda}.

\section{Simulation choices and improvements}

As in any numerical simulation, for finite temperature studies one
must makes choices of the action and parameters to be simulated.
The most important choice, for dynamical fermion simulations, is
the choice of fermion action, each of which has advantages and
drawbacks. To weigh those we recall that the
finite temperature transition or crossover is driven, for small
quark masses, by the restoration of chiral symmetry in the high
temperature phase.

\noindent (i) Wilson-type fermions, including clover fermions, have the
advantage that they are local (in fact ``ultralocal'') for any
number of fermions, while the fermion determinant is positive
for even numbers of fermions. Their big disadvantage is that the
chiral symmetry is explicitly broken and the chiral limit therefore
not protected. This makes the study of chiral symmetry restoration
cumbersome and difficult. It becomes really meaningful only in the
continuum limit.

\noindent (ii) Staggered fermions have the advantage that they have a (remnant)
chiral symmetry. The chiral limit is thus protected. In addition, they
are comparatively cheap to simulate. The main disadvantage of
staggered fermions is the need to use the ``fourth root trick'' when
the desired number of fermions of a given mass is not a multiple of four.

\noindent (iii) Overlap and domain wall fermions (at least for sufficiently
large fifth dimension $L_s$) have the advantage of good chiral symmetry
and the protection of the chiral limit, and of being local for
any number of fermions (at least for sufficiently small lattice spacing
$a$). Furthermore, for overlap fermions, the fermion determinant is
positive for any number of fermions~\cite{BHEN}. The disadvantage
of these chiral fermion discretizations is that they are expensive to
simulate. This problem is acerbated by the fact that one might need
lattices with $N_t \ge 8$ to have lattice spacings small enough in
the transition/crossover region for the fermion action to be sufficiently
local. In judging this drawback, one should keep in mind that, for the
computation of the EOS, for example, the scaling of the costs of a simulation
is worse, by a factor $\sim a^{-4}$, than for typical zero temperature
simulations, because the observable, obtained from the difference of
a finite and a zero temperature simulation, decreases, in lattice units,
as $a^4$, while the error decreases much slower.

\begin{figure}[h]
\begin{minipage}{17pc}
\includegraphics[width=16pc]{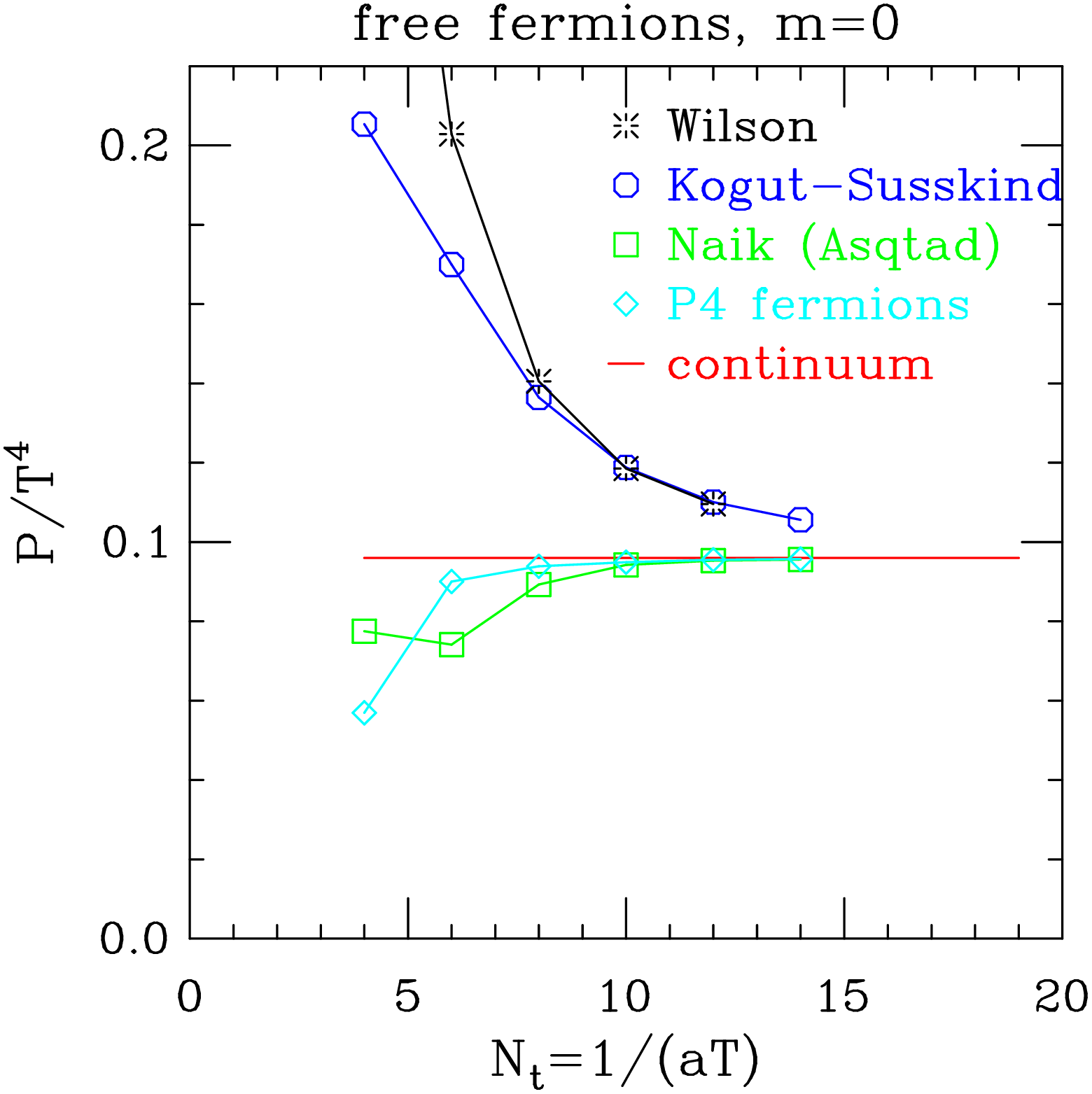}
\caption{\label{p_free} The pressure for free fermions with various fermion
discretizations versus $N_t$, from Ref.~\cite{MILC04}.}
\vspace{1.2cm}
\end{minipage}\hspace{2pc}%
\begin{minipage}{17pc}
\includegraphics[width=16pc]{del4_cmp3-lat.eps}
\caption{\label{taste_viol} The pion mass splitting, $(m^{\prime 2}_\pi -
m^2_\pi) / T^2_c$ as function of $m^2_\pi/T^2_c$ at the lattice spacing
corresponding to the finite temperature transition/crossover point,
from Ref.~\cite{AFKS05}.}
\end{minipage}
\end{figure}

Most of the recent dynamical fermion thermodynamics simulations
have used improved versions of staggered fermions. The improvement
aims to reduce taste symmetry breaking, by using some version of
``smeared'' or ``fattened'' gauge links in the nearest neighbor
hopping term. This is done for all three versions currently under
investigation, p4 (Bielefeld, RBC-Bielefeld) \cite{p4_intro},
asqtad (MILC) \cite{ASQTAD} and stout-link (Wuppertal-Budapest)
\cite{AFKS05}. The first two also aim to improve the
dispersion relation, which implies improved behavior of thermodynamic
quantities in the high temperature limit ({\it i.e.,} for
free fermions) by including three-hop terms: straight, the Naik term,
for asqtad, and ``knight moves'' for p4.
For free fermions, and hence in the high temperature limit, the
link fattening becomes inoperative. Thus stout-link fermions act
like standard staggered fermions, and asqtad fermions like Naik fermions.
The two forms of improvement are illustrated in Figs.~\ref{p_free}
and \ref{taste_viol}.

The other major advance, in the last couple of years, has come from
the algorithmic side. Because of the fractional power of the fermion
determinant, needed for the fourth-root trick, the usual, exact HMC
algorithm~\cite{HMC} can not be used. Instead, the inexact hybrid
R algorithm~\cite{R_alg}, with stepsize errors of ${\cal O}(\epsilon^2)$,
was employed. This obstacle was recently overcome with the invention
of the exact RHMC algorithm~\cite{RHMC}. As discussed in the
talk by M.~Clark~\cite{RHMC_lat06} the RHMC algorithm is not only
exact but allows various other improvements -- multiple time-step
integration schemes, multiple pseudofermion fields, etc. -- that can
speed up the simulations by factors of 2 up to 8 compared to the
R algorithm.

A couple of exploratory studies with Wilson-type fermions were presented
at this conference, using a hypercube action~\cite{Shcheredin} and
with twisted-mass Wilson fermions~\cite{Lombardo}.
But in the rest of this talk I will concentrate on simulations with
(improved) staggered fermions. For the purpose of this talk the
validity of the fourth-root trick, discussed in detail by S.~Sharpe
\cite{4_th_root}, will be assumed.

\section{The phase diagram}

\subsection{The physical point}

\begin{figure}[h]
\centering
\includegraphics[width=12cm]{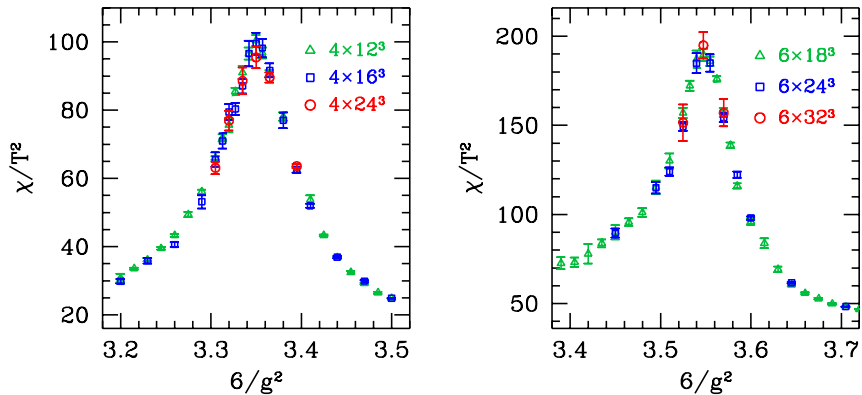}
\caption{\label{chi_4_6_fodor} The chiral susceptibility for $N_t=4$ (left)
and $N_t=6$ (right) and various volumes, from Ref.~\cite{Szabo}.}
\end{figure}

Determining the nature of the finite temperature transition or crossover
at the physical point is not only of interest in its own right, but
has implications on the possible phase diagram with finite chemical
potential (see {\it e.g.} the talks by C.~Schmidt~\cite{Schmidt} and
M.~Stephanov\cite{Stephanov}). The Wuppertal-Budapest group \cite{Szabo}
made a systematic investigation using stout-link improved fermions
and the exact RHMC algorithm. They used a physical strange quark
mass and physical (degenerate) light quark masses.
The chiral susceptibility for $N_t=4$ and $6$ and various volumes is
shown in Fig.~\ref{chi_4_6_fodor}. It shows no sign of increasing with
volume as would be expected for a genuine phase transition, indicating
existence of only a crossover.

\begin{figure}[h]
\centering
\includegraphics[width=14cm]{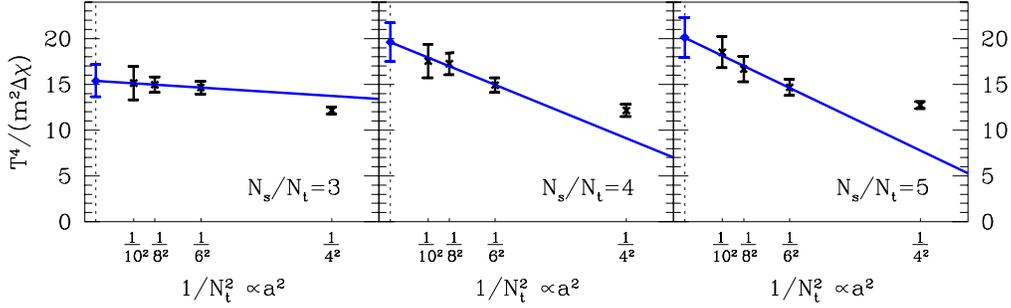}
\caption{\label{chi_cont_fodor} The continuum extrapolation of $T^4 /
(m_q^2 \Delta \chi)$ for fixed aspect ratio $N_s/N_t = 3, 4$ and $5$,
from Ref.~\cite{Szabo}.}
\end{figure}

A definite statement, however, needs an extrapolation to the continuum
and infinite volume limits. For a meaningful extrapolation renormalized
observables have to be considered. The Wuppertal-Budapest group does
this by subtracting the $T=0$ value, $\Delta \chi = \chi(T) - \chi(T=0)$,
to cancel a potential additive divergence, and then multiplies with
$m_q^2$ to obtain an RG invariant observable. This is then extrapolated
to the continuum limit for fixed aspect ratio $N_s/N_t = 3, 4$ and $5$,
see Fig.~\ref{chi_cont_fodor}.

The resulting continuum renormalized susceptibility stays finite
and non-zero in the infinite volume limit. It would diverge for a
genuine phase transition. Hence we now have convincing evidence that,
in nature, QCD has a crossover at finite temperature and zero baryon
chemical potential.

\subsection{The second order boundary line}

De~Forcrand and Philipsen mapped out the second order critical line
that separates the first order region at small quark masses from the
crossover region at intermediate quark masses in the $m_{ud} - m_s$ plane
of the phase diagram~\cite{dFP06}. They used standard, {\it i.e.,}
unimproved, staggered fermions and the exact RHMC algorithm.

\begin{figure}[h]
\centering
\includegraphics[width=9cm]{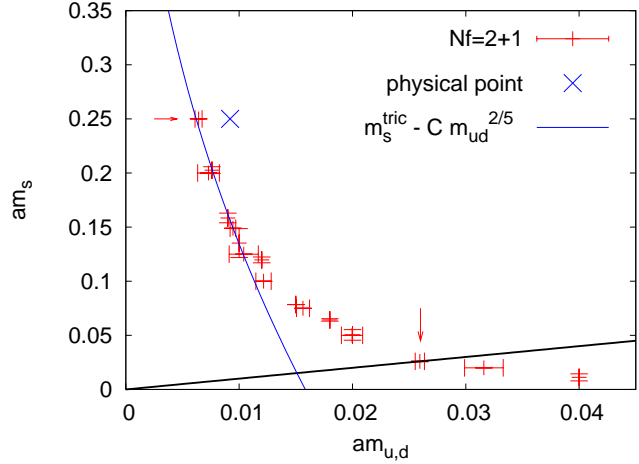}
\caption{\label{bound_line_deF} Determination of the second order
boundary line between the first order and crossover regions in the
$m_{ud} - m_s$ plane, from Ref.~\cite{dFP06}.}
\end{figure}

At fixed $am_s$ they determined the location of the second order point,
as function of $am_{ud}$, by requiring that the Binder cummulant,
$B_4(m_{ud}^{crit})$ attain its critical Ising value, $\simeq 1.6$.
The locations, with errors, are shown in Fig.~\ref{bound_line_deF}.
The arrows in the plot indicate parameters values at which
zero-temperature simulations were performed to determine meson masses
and lattice spacing. De~Forcrand and Philipsen found
\begin{itemize}
\item For the physical strange quark mass, the second order boundary
occurs at non-vanishing $m_{ud}^{crit} < m_{ud}^{phys}$, {\it i.e.,}
the physical point is in the crossover region, in agreement with the
result of the Wuppertal-Budapest group.
\item The critical line is consistent with $m_s^{crit}(m_{ud}) =
m_s^{tric} - c m_{ud}^{2/5}$, {\it i.e.,} the behavior that is expected
if a tricritical point exists.
\end{itemize}

One should emphasize that these results are most likely qualitative,
only. Unimproved staggered fermions were used on lattices with $N_t=4$,
were the lattice spacing in the crossover/transition region is quite
large, $a \sim 0.27$ fm. Therefore lattice effects could be significant.
For example, at the second order critical point for three degenerate
flavors, de~Forcrand and Philipsen find $m_\pi/m_\rho = 0.304(2)$,
whereas the Bielefeld group, using p4 fermions, concluded that
\cite{p4_nf3} $m_\pi/m_\rho < 0.18$, the physical value of the mass ratio.
For additional results with three degenerate flavors see also the
contribution by Cheng~\cite{Cheng}.

\subsection{Massless 2 flavor $\chi$QCD}

Kogut and Sinclair continued their study of massless $N_f=2$ $\chi$QCD
on $N_t=8$ lattices with $N_s=12, 16$ and $24$~\cite{KS06}. $\chi$QCD
contains an irrelevant chiral 4-fermion interaction, which makes the
fermion matrix non-singular in the limit of zero (bare) quark mass
while preserving the chiral symmetry of the Lagrangian,
thus enabling the study of spontaneous chiral symmetry breaking
in the massless limit. For numerical simulations, the 4-fermion
interaction is made quadratic in the fermions by introducing an
auxiliary $(\sigma,\pi)$ field.

\begin{figure}[h]
\begin{minipage}{17pc}
\includegraphics[width=15pc]{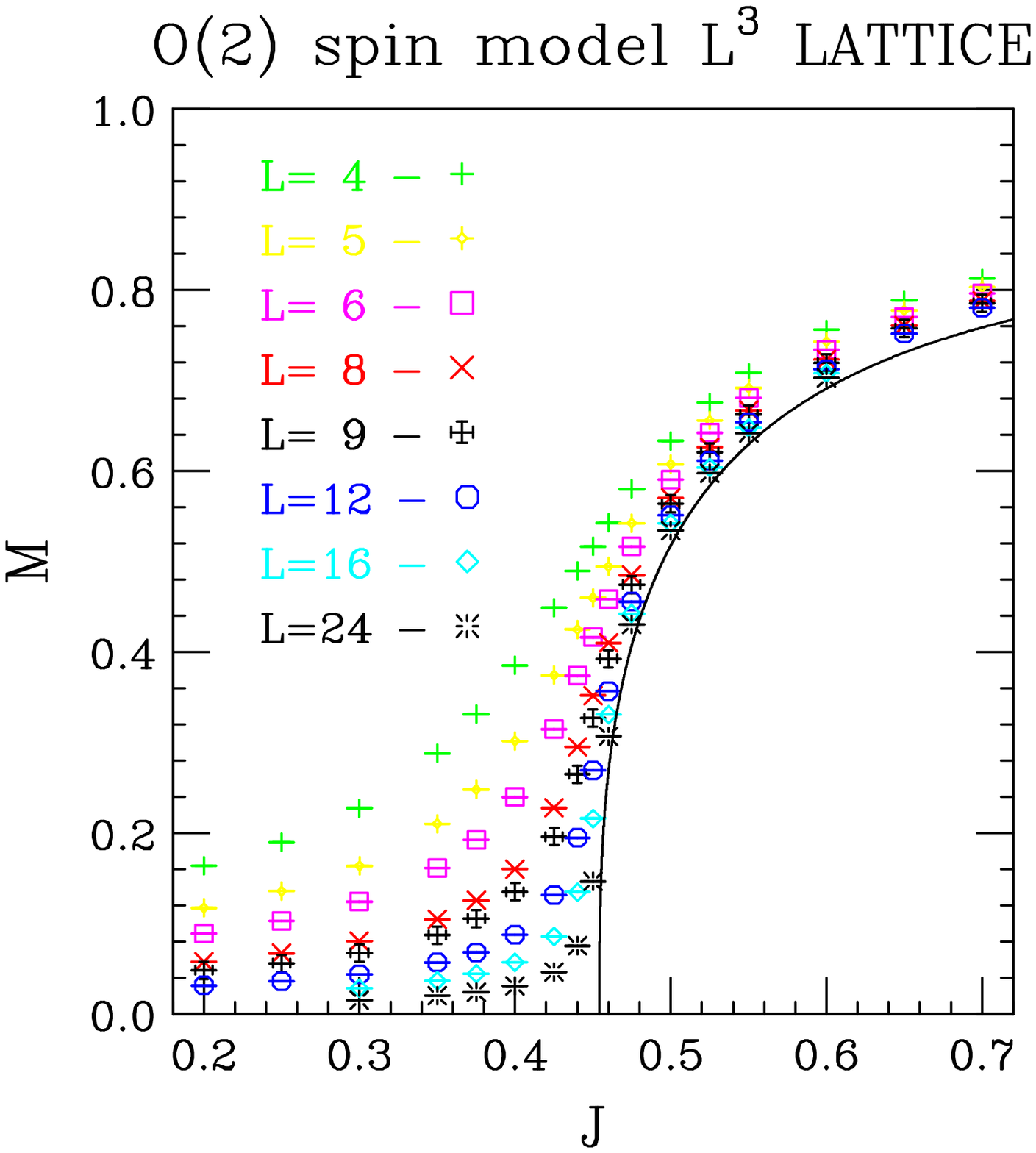}
\end{minipage}\hspace{2pc}%
\begin{minipage}{17pc}
\includegraphics[width=16pc]{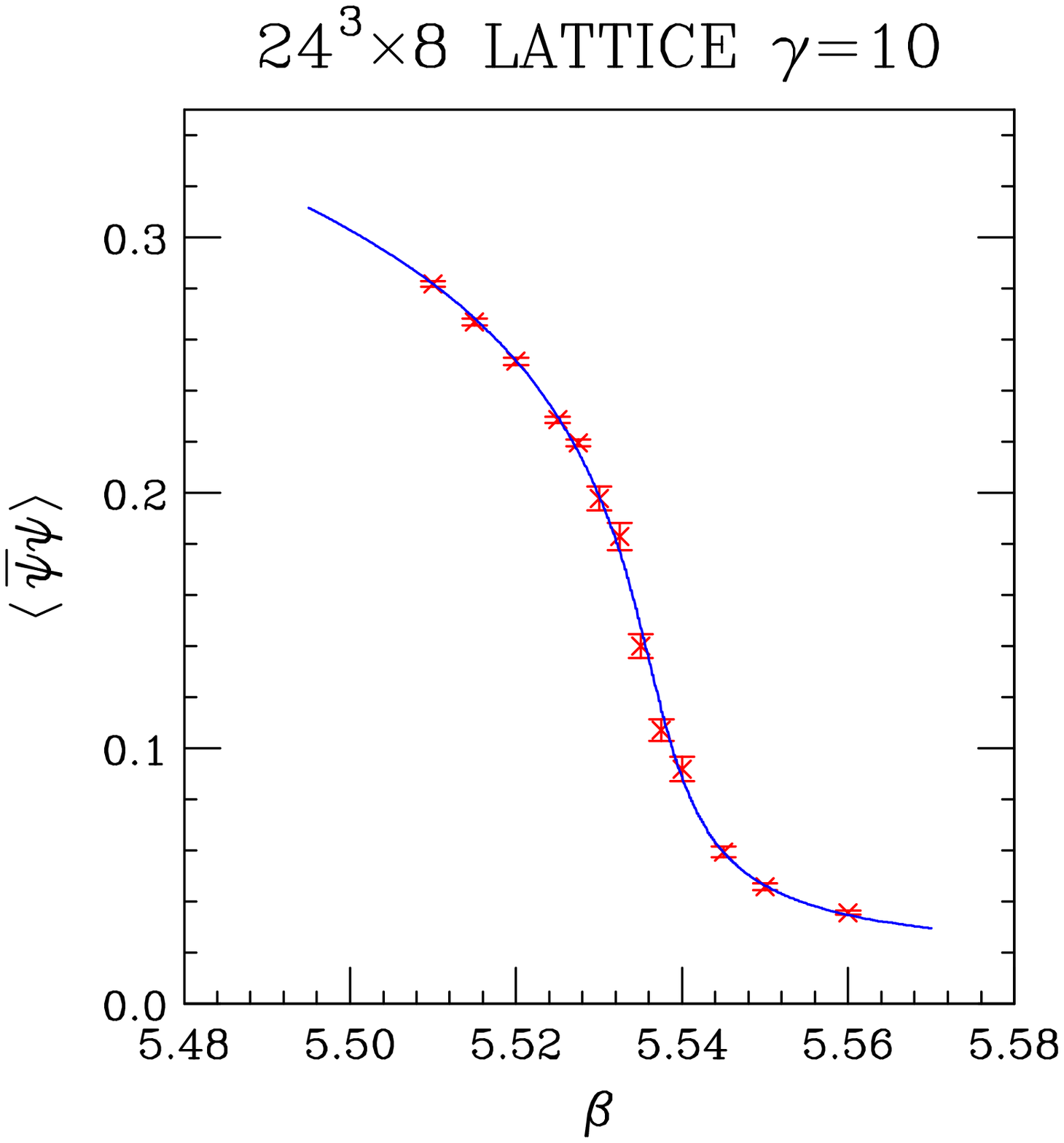}
\end{minipage}
\caption{\label{O2_M_pbp} The magnetization of the $O(2)$ spin model
in the transition region for various volumes (left) and an example of
a best volume fit of $\langle \bar \psi \psi \rangle$ to the $O(2)$ spin
model magnetization (right), from Ref.~\cite{KS06}.}
\end{figure}

Kogut and Sinclair used (standard) staggered fermions which, at finite
lattice spacing have a $U(1) \times U(1)$ chiral symmetry, expected
to be spontaneously broken to $U(1)$. Hence, if the chiral symmetry
restoration transition is second order, one expects the transition
to be in the 3-d $O(2)$ universality class.
Therefore, one wants to compare to the 3-d $O(2)$ spin model.
However, the magnetization of the $O(2)$ spin model has large finite
size effects, as seen in Fig.~\ref{O2_M_pbp} (left). Hence, instead
of comparing to infinite volume $O(2)$ behavior, Kogut and Sinclair
compare to finite volume behavior for a ``best matched size''.
The strategy, thus, is to find the volume for which the $O(2)$
magnetization gives the best fit
\begin{equation}
\langle \bar \psi \psi (\beta) \rangle =
b \; \langle M( a(\beta-\beta_c) + T_c) \rangle \qquad \quad
i.e. \quad 1/J = T = a(\beta-\beta_c) + T_c ~.
\end{equation}
of the chiral condensate, for a given QCD lattice size.

Such fits work well, see Fig.~\ref{O2_M_pbp} (right),
strongly suggesting that the $\chi$QCD data are
compatible with the $O(2)$ spin model. The corresponding $O(2)$
volumes are small,  $6^3$ for $16^3 \times 8$ and $8^3$ for
$24^3 \times 8$. Thus, one should not attempt comparisons with
large volume critical behavior.
Kogut and Sinclair find, in retrospect, that such fits also work for
their earlier $N_t=6$ data. Hence the conclusions of their earlier
work, Ref.~\cite{KS00-01}, should be modified accordingly.

\subsection{2 flavor LQCD with KS quarks at $N_t=4$}

The Pisa group~\cite{Pica} continued their investigation of the phase
transition of $N_f=2$ QCD with standard staggered quarks at $N_t=4$.
To check their previous results~\cite{Pisa05}, obtained with the inexact
R-algorithm they performed comparisons with the exact RHMC algorithm.
They found that the systematic ${\cal O}(\epsilon^2)$ step-size errors
are comparable to the statistical ones.

\begin{figure}[h]
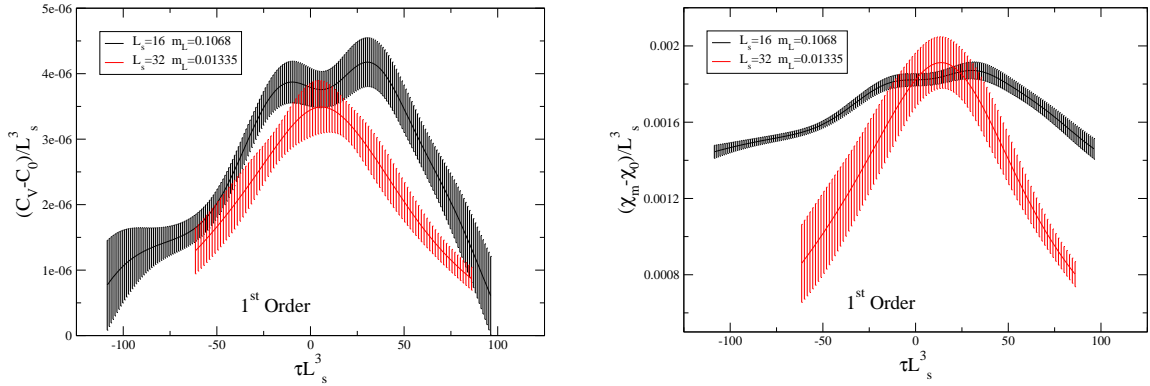

\vspace{4mm}
\begin{minipage}{17pc}
\includegraphics[width=17pc]{susc_scal_run3.eps}
\end{minipage}\hspace{2pc}%
\begin{minipage}{17pc}
\includegraphics[width=17pc]{susc_chi_scal_run3.eps}
\end{minipage}
\caption{\label{1st_ord_scal} Test of first order finite volume scaling
with $m_L L^3$ kept constant, from Ref.~\cite{Pica}.}
\end{figure}

They then tested their hypothesis of a first order transition with a
finite size scaling study of the specific heat, $C_V$, and chiral
susceptibility, $\chi_m$, with the scaling variable $m_L L^{y_h}$ kept
constant for the choice $y_h=3$, the value for a first order transition,
see Fig.~\ref{1st_ord_scal}. The specific heat scales nicely, but not
the chiral susceptibility. They speculate that the reason for the latter
might be the large mass needed for the smaller volume,
$am_{L=16} = 0.1068$.

\begin{figure}[h]
\centering
\includegraphics[width=9cm]{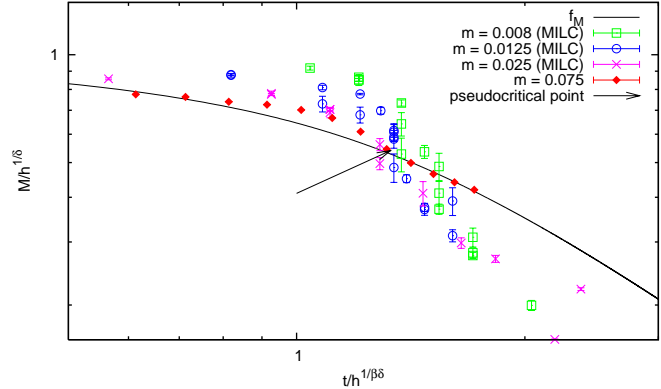}
\caption{\label{O4_scal} Comparison of $N_f=2$ QCD data with staggered
fermions at $N_t=4$ to the $O(4)$ scaling function, from
Ref.~\cite{Mendes05}.}
\end{figure}

On the other hand, T.~Mendes~\cite{Mendes05} recently compared previous
data by the MILC collaboration~\cite{MILC99} and her own new data at
a heavier mass, $am = 0.075$, with the (infinite volume) $O(4)$ scaling
function, as shown in Fig.~\ref{O4_scal}. Mendes inferred that the scaling
works quite well, especially at the heavier mass. 

I would conclude that the issue of the order of the phase transition
for 2-flavor QCD, even with staggered fermions at $N_t=4$, and certainly
in the continuum limit, remains an open question.

\subsection{$T_c$ for full QCD}

The RBC-Bielefeld collaboration~\cite{p4_Tc21,Umeda} has recently performed
a systematic study of the crossover temperature in full QCD, {\it i.e.},
physical strange quark mass, and degenerate light up and down quark
masses, $m_l$. They used p4 fermions and the exact RHMC algorithm for
simulations with $N_t=4$ and $6$, several light quark masses $m_l$,
and fairly large volumes, $2 \le N_s/N_t \le 4$.

\begin{figure}[h]
\begin{minipage}{17pc}
\includegraphics[width=17pc]{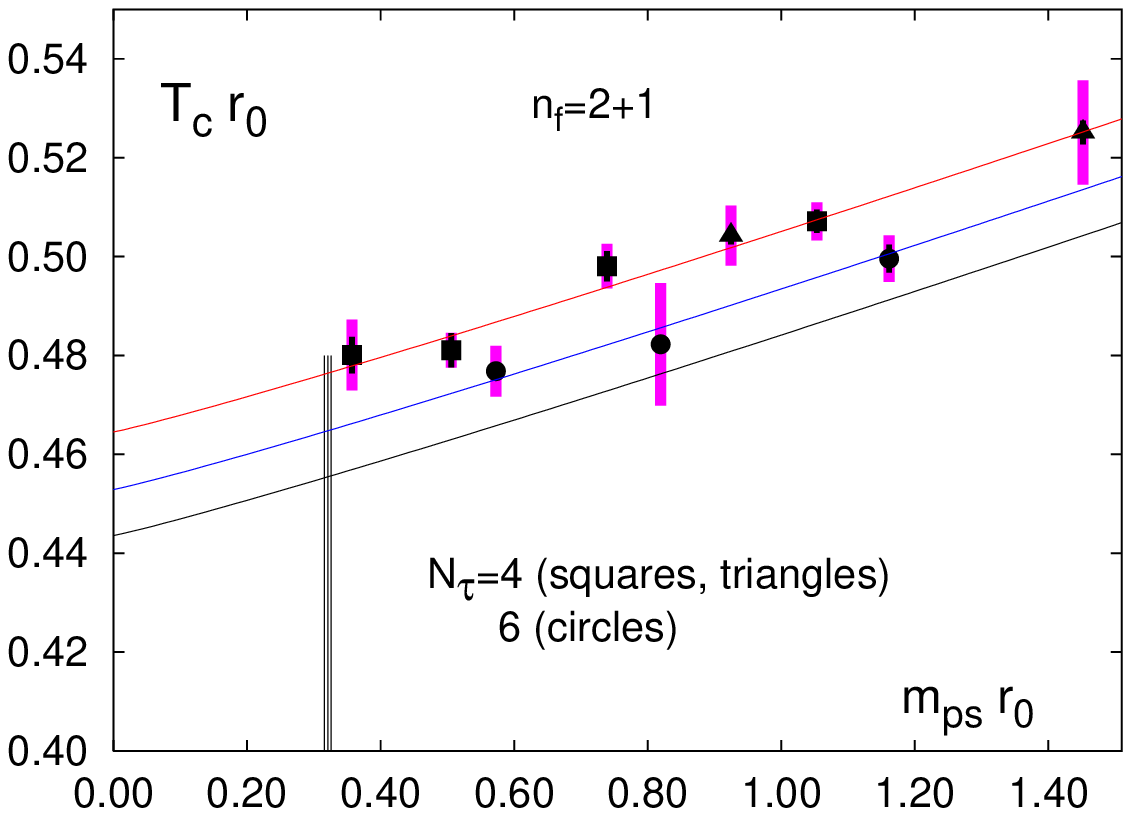}
\caption{\label{Tc_f21_p4} $T_c$ in units of $r_0^{-1}$
as function of $m_{PS} r_0$, from Ref.~\cite{p4_Tc21}.
The vertical line shows the location of the physical value.}
\end{minipage}\hspace{2pc}%
\begin{minipage}{17pc}
\vspace{-1pc}
\hspace{-1pc}
\includegraphics[width=18pc]{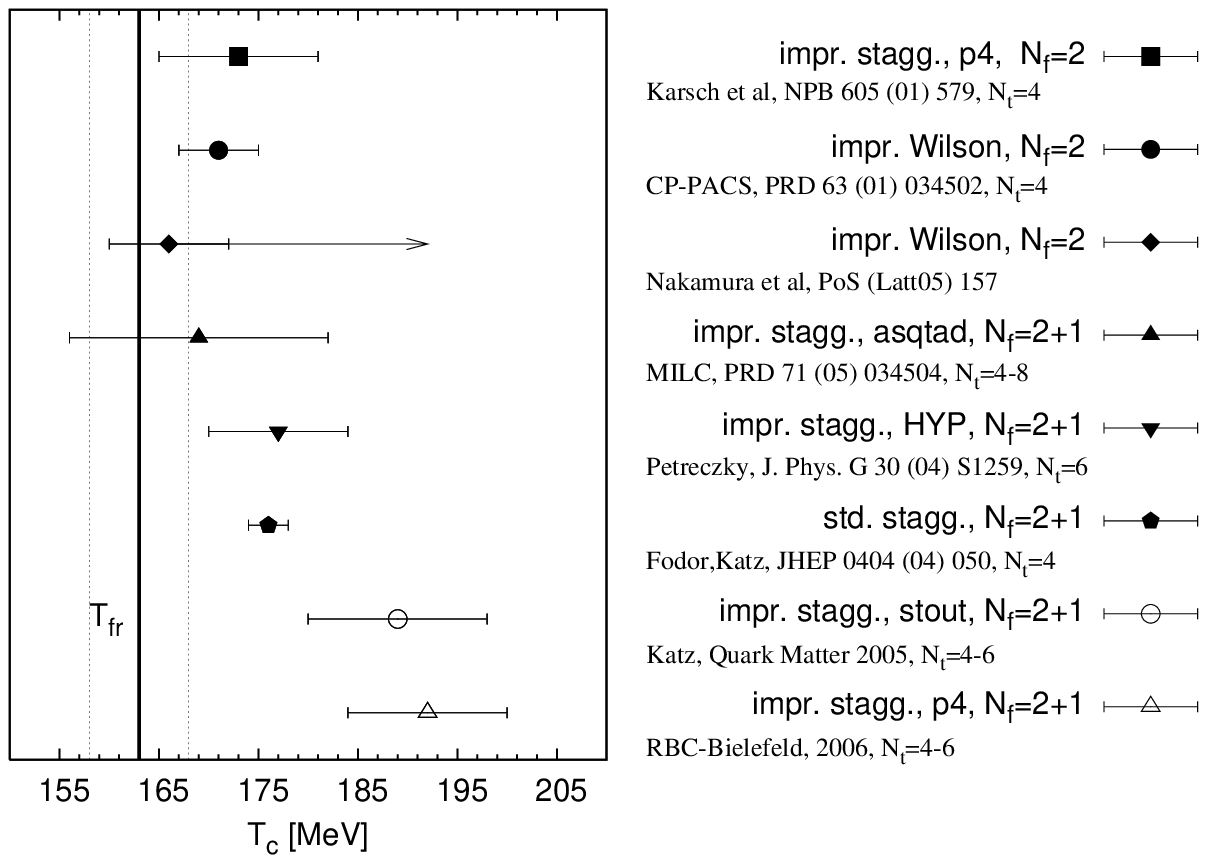}
\caption{\label{Tc_all} Recent lattice determinations of $T_c$,
collected by P.~Petreczky in Ref.~\cite{Tc_coll}.}
\end{minipage}
\end{figure}

For each simulation parameter set, they located the crossover point
from the peak position of the chiral and Polyakov loop susceptibilities.
The two peak positions appeared to coincide at large volume, suggesting
that the crossovers for deconfinement and chiral symmetry restoration
occur simultaneously. So the authors averaged over the two determinations
at finite volume. $T_c$, in units of $r_0^{-1}$ is shown in
Fig.~\ref{Tc_f21_p4} as function of $m_{PS} r_0$. The data at finite
lattice spacing, $a \propto 1/N_t$, and unphysical light quark masses are
then extrapolated to the continuum limit and physical point, given by
$m_{PS} r_0 = 0.321(5)$ --- $r_0$ was taken from Ref.~\cite{Gray05}:
$r_0 = 0.469(7)$ fm --- using the form
\begin{equation}
(T_c r_0)_{m_l,N_t} = T_c r_0 + A (m_{PS} r_0)^d + B/N_t^2 ~.
\label{eq:Tc_f21_extr}
\end{equation}
Here, $d = 1.08$ would be expected for a second order phase transition
(in the $O(4)$ universality class) and $d = 2$, {\it i.e.}, a linear
dependence on $m_l$, for a first order transition. At the physical
point they find $T_c r_0 = 0.457(7)^{+8}_{-3}$, with the second error
coming from using $d=2$ and $d=1$ in the extrapolation.

In physical units, the transition/crossover temperature is
$T_c = 192(7)(4) \; \MeV\ $. This is about 12\% larger than the value
obtained by the MILC collaboration, using asqtad fermions, the
R-algorithm, lattices with $N_t=4, 6, 8$ and $N_s/N_t=2$, and
a combined chiral/continuum limit extrapolation as in
eq.~(\ref{eq:Tc_f21_extr}). The MILC collaboration obtained
$T_c = 169(10)(4) \; \MeV\ $~\cite{MILC04}. While they worked in units
of the scale $r_1$, the conversion to physical units was done with
compatible values and does not contribute to the disagreement.
A collection of recent determinations of $T_c$, assembled by
P.~Petreczky in Ref.~\cite{Tc_coll}, is shown in Fig.~ \ref{Tc_all}.

\section{The equation of state}

After establishing the phase diagram, including the order and nature of
the phase transitions and/or crossovers, one would like to understand
the nature of the different phases. One of the basic quantities for
this is the equation of state (EOS), the pressure $p(T)$, the entropy
density $s(T) = dp(T)/dT$ and the energy density $\epsilon(T) =
T s(T) - p(T)$. Besides its intrinsic interest as a fundamental
property of QCD, the EOS is of phenomenological interest. For example,
it is an import input in hydrodynamical models of the QGP, often
used to try to interpret results from heavy-ion collision experiments,
such as the observed elliptic flow. For a quantitative understanding,
obviously, a quantitative understanding of the EOS is necessary.

\subsection{Low temperature behavior}

\begin{figure}[h]
\centering
\includegraphics[width=5cm,angle=270]{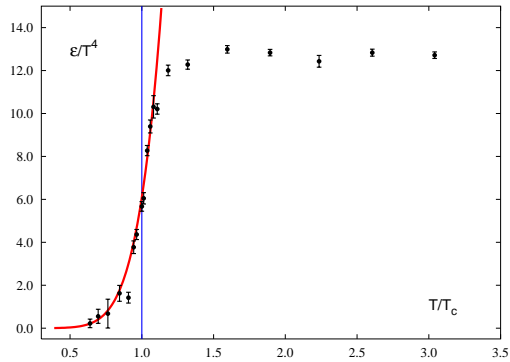}
\caption{\label{HRG_comp} Comparison of the EOS from lattice simulations
with a hadron resonance gas model with masses adjusted to the lattice quark
masses used, from Ref.~\cite{KRT03}.}
\end{figure}

In the low temperature hadronic phase, the excitations are weakly
interacting hadrons, including all resonances. This suggests that a
hadron resonance gas model (HRG) should give a good description of
the EOS. This was recently tested by a comparison with lattice
data using p4-fermions and $N_t=4$~\cite{KRT03}. Since the light
quark mass used was larger than the physical light quark mass, the
hadron masses of the HRG where adjusted to match those on the lattice.
With this adjustment the HRG works surprisingly well, even around the
crossover temperature, as can be seen in Fig.~\ref{HRG_comp}.

\subsection{The EOS at high temperatures}

At (very) high temperatures the running coupling $g(T)$ vanishes
(logarithmically in $T$), and one expects that the EOS should be
computable in perturbation theory. PT, though, is afflicted with infrared
divergences, requiring resummation of certain classes of diagrams,
giving {\it e.g.} an ${\cal O}(g^3)$ and an ${\cal O}(g^5)$ contribution.
At ${\cal O}(g^6)$ the IR divergences cause all orders of PT to
contribute, so that this order needs to be computed non-perturbatively.
The last perturbative order, $g^6 \log(1/g)$ has recently been computed
using the technique of dimensional reduction~\cite{KLRS03}.
 
\begin{figure}[h]
\begin{minipage}{17pc}
\includegraphics[width=15pc]{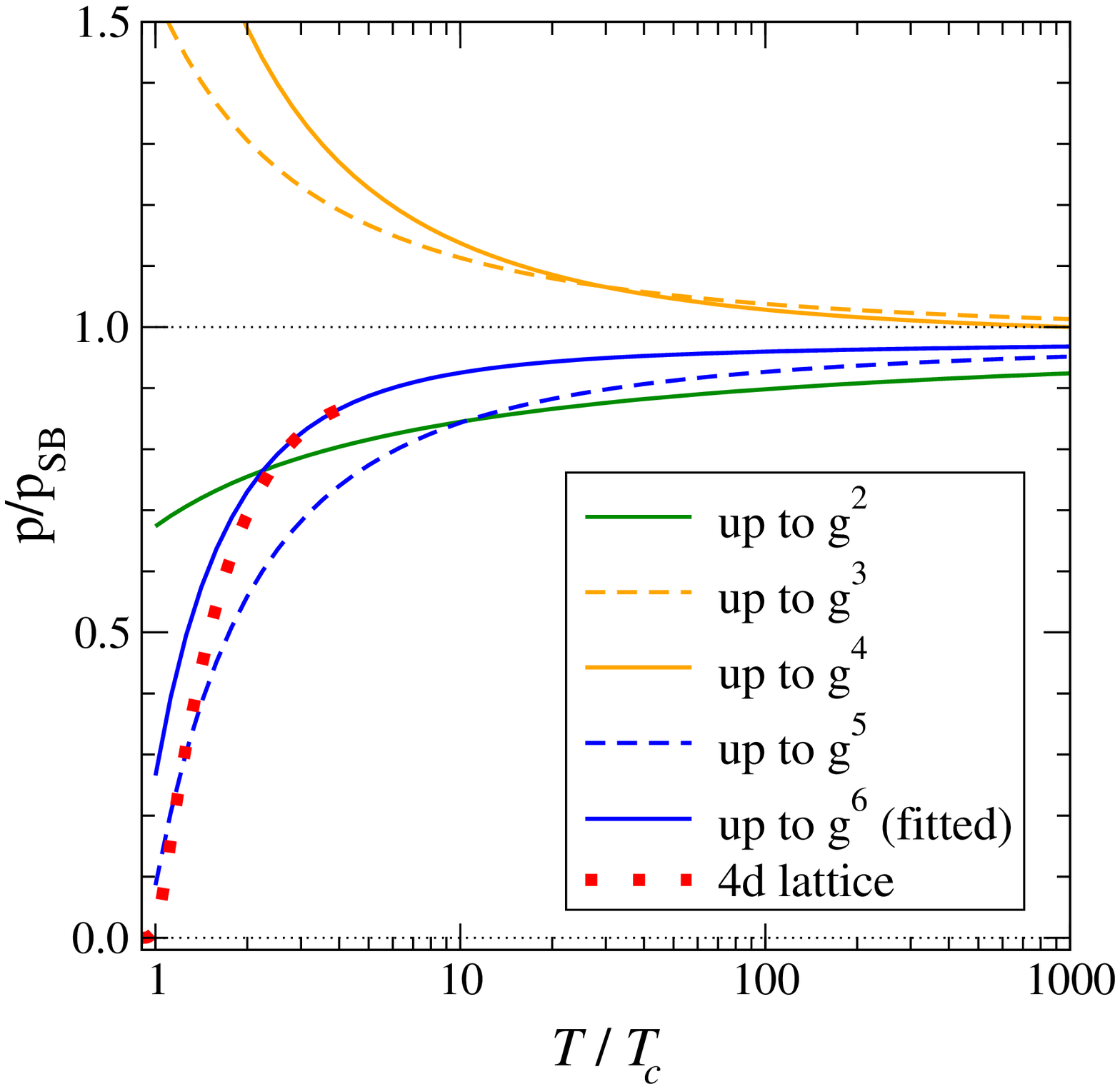}
\end{minipage}\hspace{2pc}%
\begin{minipage}{17pc}
\includegraphics[width=15pc]{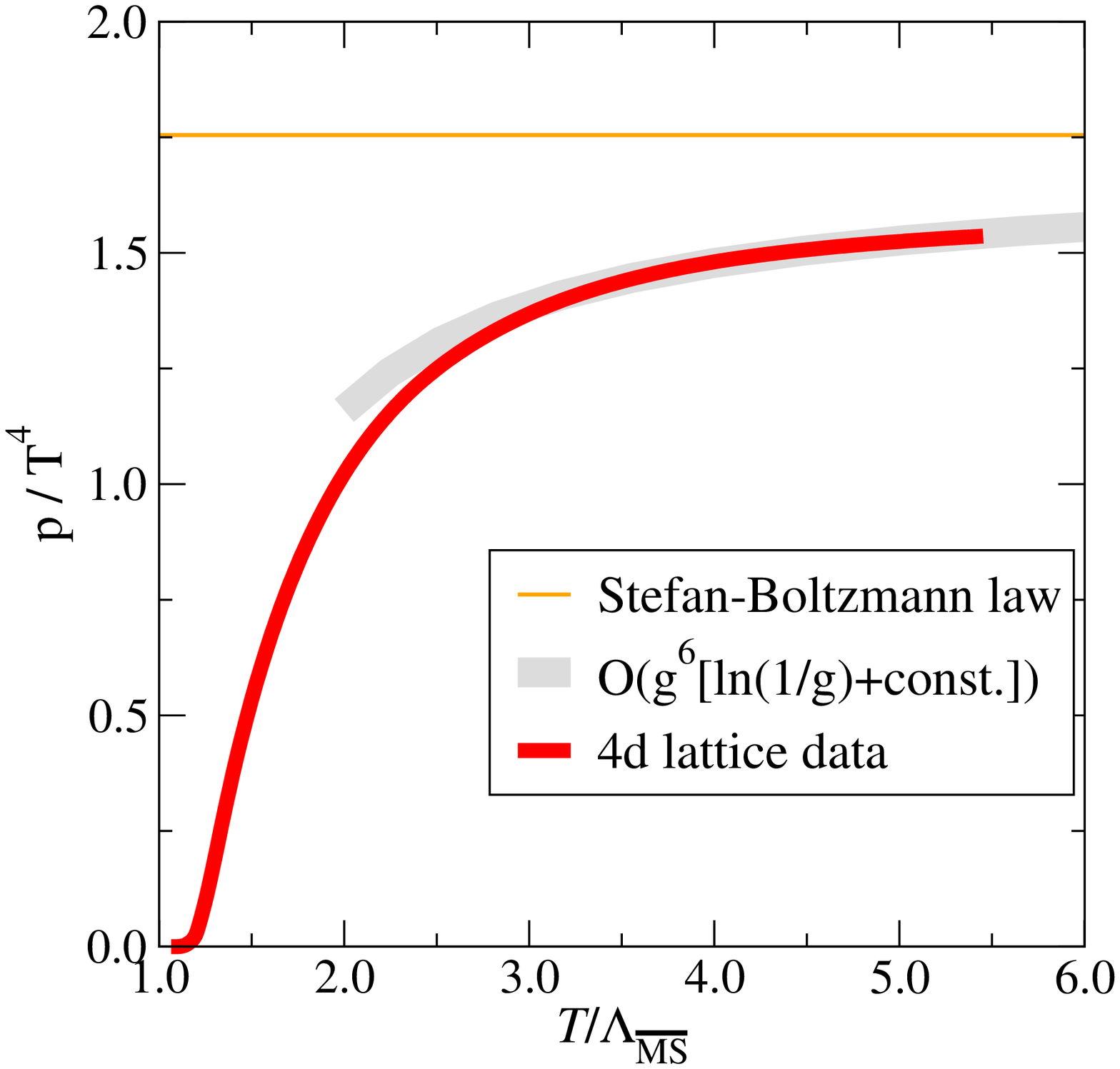}
\end{minipage}
\caption{\label{EOS_highT_comp} The pure gauge EOS in perturbation
theory and the comparison with LGT data, from Refs.~\cite{KLRS03,LS06}.}
\end{figure}

The convergence, for pure gauge theory, is illustrated in
Fig.~\ref{EOS_highT_comp}. While the convergence is questionable,
with fitting the unknown, non-perturbative $c_6 g^6$ term to (pure gauge)
LQCD results for $T > 3T_c$, good agreement can be found~\cite{LS06}.
The EOS, at high temperature, is thus reasonably well understood.
The approach to the Stefan-Boltzmann (SB) limit is rather slow, due to
the slow, only logarithmic vanishing of $g(T)$ with increasing
temperature.

The high temperature behavior of pure gauge LGT in $2+1$ dimensions
was recently investigated in detail~\cite{BDMP06}. In $2+1$-$d$, the
dimensionless ratio $g^2/T$ serves as the running coupling that
vanishes at high temperatures. Since the interaction measure (the
trace anomaly) vanishes for free gluons, one expects
\begin{equation}
\frac{\epsilon - 2p}{T^3} \simeq \frac{g^2}{T} ~ f \left( \log \frac{T}
{g^2} \right) ~.
\label{eq:I_2+1d}
\end{equation}

\begin{figure}[h]
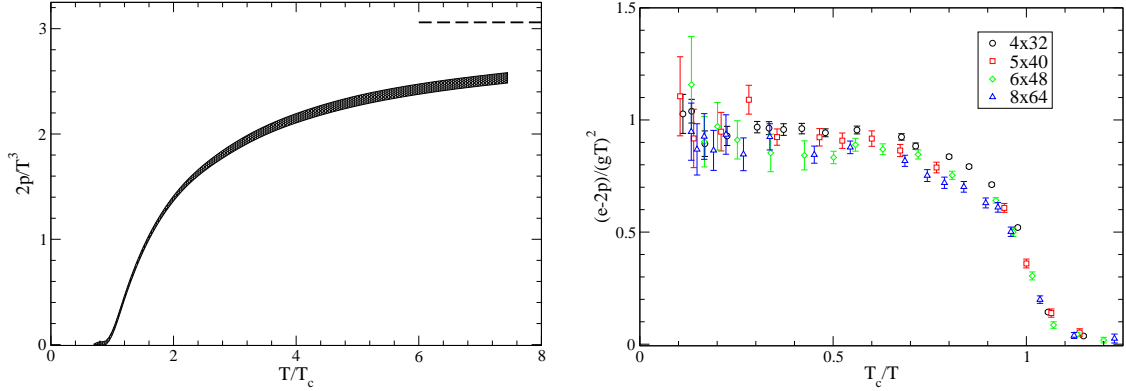

\begin{minipage}{17pc}
\includegraphics[width=17pc]{press.inf.4+6+8.eps}
\end{minipage}\hspace{2pc}%
\begin{minipage}{17pc}
\hspace{-1pc}
\includegraphics[width=17pc]{em2p.scal.all.luki.eps}
\end{minipage}
\caption{\label{EOS_2+1d} The pressure (left) and the interaction measure,
scaled to leading perturbative behavior, (right) for purge gauge LGT
in $2+1$ dimensions, from Ref.~\cite{BDMP06}.}
\end{figure}

Fig.~\ref{EOS_2+1d} shows pressure and interaction measure, scaled to
the leading perturbative behavior. The data nicely confirm that
$(\epsilon-2p)/(gT)^2$ approaches a constant at high temperature.
The EOS is, again, well understood at high temperature, increasing
our confidence in out understanding of the $3+1$-$D$ theory. Due to the
faster running of the coupling, the pressure around $T \sim 2T_c$ is
even farther from the SB limit than for $3+1$-$d$ QCD.

\subsection{The EOS around the transition/crossover region}

While the EOS is well understood and modeled at low and high temperatures
the non-perturba\-tive input from lattice calculations is needed to
determine the EOS at intermediate temperatures, {\it i.e.}, in the
transition/crossover region. It is also needed, as we have seen, to calibrate
the high temperature description --- via the fit of the ${\cal O}(g^6)$
term --- and to check the range of validity of the HRG model at low
temperatures.

\begin{figure}[h]
\begin{minipage}{17pc}
\includegraphics[width=15pc]{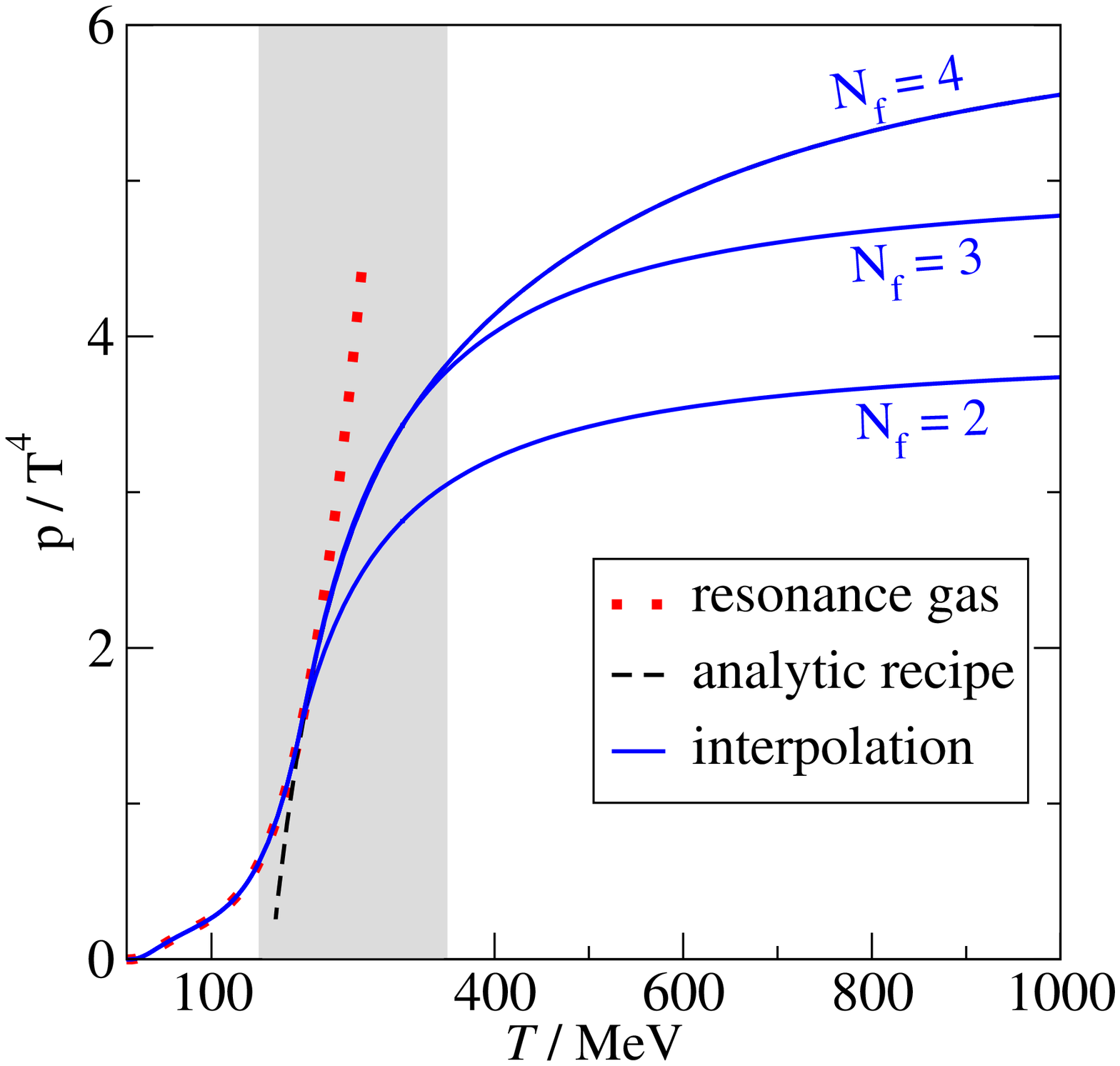}
\caption{\label{EOS_highT_qmass} Quark mass effects from NLO perturbation
theory on a ``phenomenological modeling'' of the EOS~\cite{LS06} (see text).}
\end{minipage}\hspace{2pc}%
\begin{minipage}{17pc}
\vspace{-4mm}
\hspace{-2pc}
\includegraphics[width=19pc]{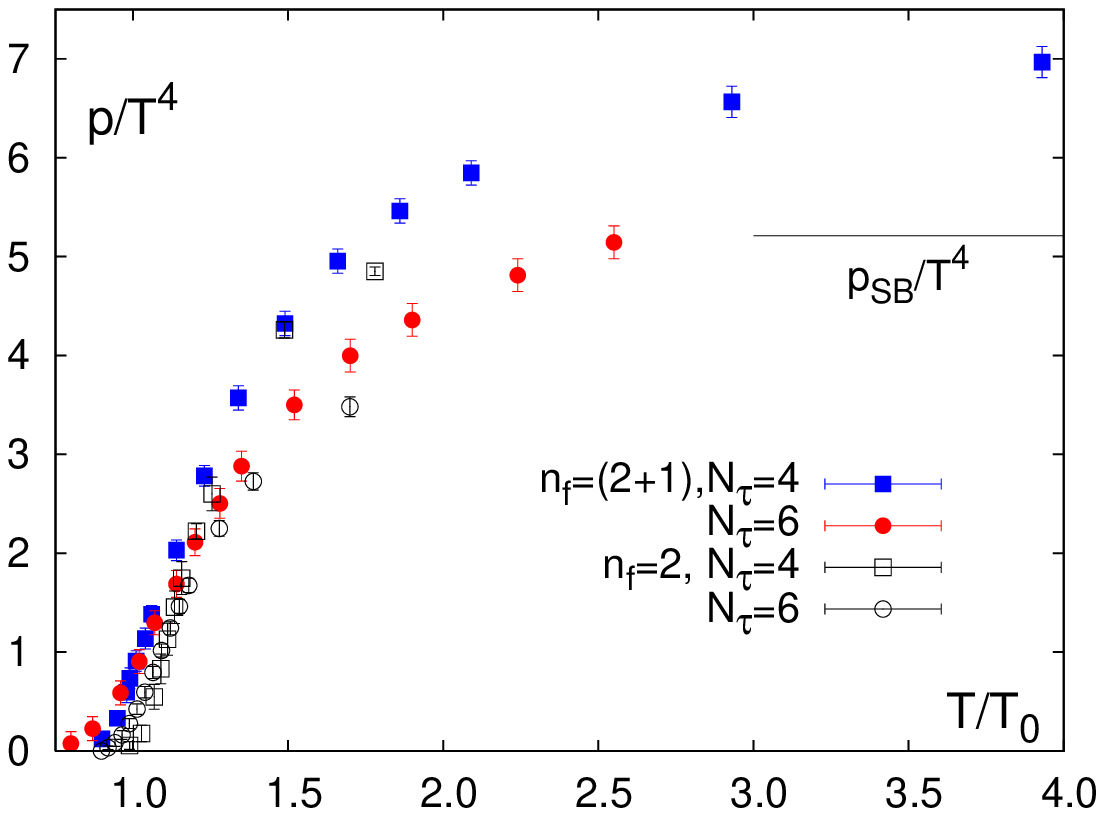}
\caption{\label{EOS_LGT_qmass} Quark mass effects in the lattice EOS
with staggered quarks. $N_f=2$ standard staggered data are from
Refs.~\cite{old_MILC_Nt4,old_MILC_Nt6} and $N_f=2+1$ stout-link data
from Ref.~\cite{AFKS05}.}
\end{minipage}
\end{figure}

This intermediate temperature range is also the region where the effect
of the strange quark, with $m_s/T_c \sim {\cal O}(1)$, and even of the
charm quark, become visible. This can be seen in Fig.~\ref{EOS_highT_qmass}
which shows the quark mass effects incorporated as NLO perturbative
corrections to a phenomenological modeling of the EOS based on the
HRG, pure gauge lattice results and perturbation theory~\cite{LS06},
and Fig.~\ref{EOS_LGT_qmass}, which compares lattice results for
staggered quarks with $N_f=2$~\cite{old_MILC_Nt4,old_MILC_Nt6} and
$N_f=2+1$~\cite{AFKS05} flavors\footnote{Incidentally, this figure
also indicates that the stout-link improvement has negligible impact
for the EOS.}. The effect of the strange quark becomes clearly noticeable
around $T \approx 1.5 T_c$ and the effect of the charm quark above
$T \sim 3 T_c$.

\begin{figure}[h]
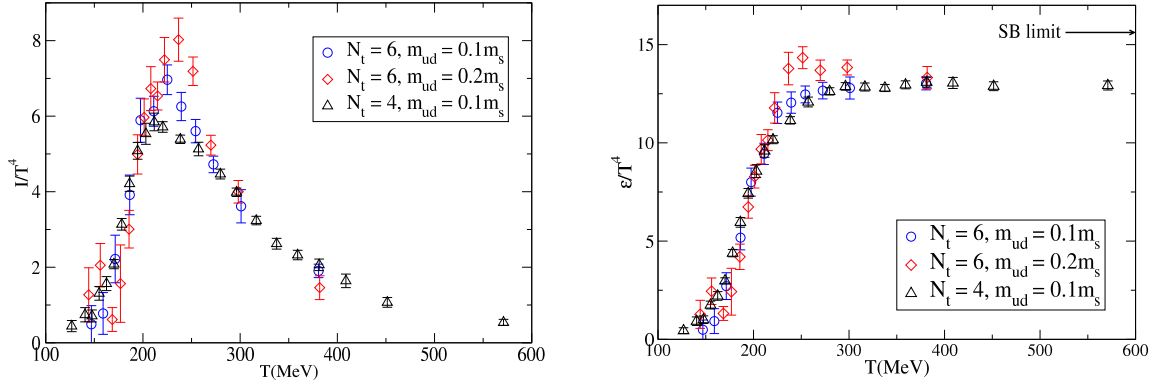

\begin{minipage}{17pc}
\includegraphics[width=17pc]{I_pbp_corr.eps}
\end{minipage}\hspace{2pc}%
\begin{minipage}{17pc}
\includegraphics[width=17pc]{E_pbp_corr.eps}
\end{minipage}
\caption{\label{EOS_MILC} The interaction measure (left) and the energy
density (right) from $2+1$ flavor simulations with asqtad fermions
along two lines of constant physics~\cite{Levkova}.}
\end{figure}

\begin{figure}[h]
\begin{minipage}{17pc}
\includegraphics[width=17pc]{P_pbp_corr.eps}
\end{minipage}\hspace{2pc}%
\begin{minipage}{17pc}
\vspace{-5mm}
\hspace{-1pc}
\includegraphics[width=18pc]{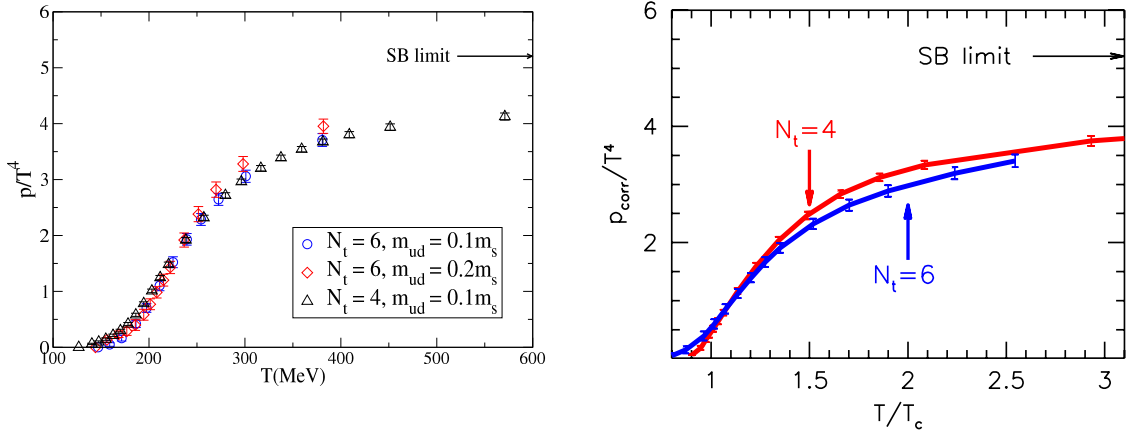}
\end{minipage}
\caption{\label{EOS_2+1f} The pressure with asqtad fermions, from
Ref.~\cite{Levkova}, (left) and with stout-link fermions, corrected for
tree-level lattice effects, from Ref.~\cite{AFKS05}, (right) for full
QCD simulations.}
\end{figure}

There have been two recent computations of the EOS in full QCD, on the
path to a determination in the continuum limit, both using $N_t=4$ and
$6$ lattices. The first, using stout-link improved fermions and the exact 
RHMC algorithm, appeared in Ref.~\cite{AFKS05}. Since stout-link fermions
do not have an improved high temperature behavior, the authors of
Ref.~\cite{AFKS05} applied an a-posteriori tree-level correction factor
$c_{cont}/c_{N_t} = 0.571$ and $0.663$ for $N_t=4$ and $6$ to correct
for the tree level ${\cal O}(a^2)$ lattice effects (see difference in
the data presented in Fig.~\ref{EOS_LGT_qmass} and Fig.~\ref{EOS_2+1f}
(right)).

The second, by the MILC collaboration, described in more detail in the
contribution by L.~Levkova~\cite{Levkova}, uses asqtad fermions. For the
most part, configurations generated earlier with the
R-algorithm~\cite{MILC04} were analyzed. The major effort, over the last
six months, concerned an estimate of, and correction for, the step-size
errors induced by the R-algorithm as described in more detail in the
contribution by L.~Levkova~\cite{Levkova}.

The preliminary, step-size corrected results along two different lines
of constant physics, with $m_l = 0.2 m_s$ and $m_l = 0.1 m_s$, and two
$N_t$'s, $4$ and $6$ for the latter case, for the interaction measure
and the energy density are shown in Fig.~\ref{EOS_MILC}. Overall, light
quark mass and lattice spacing effects appear to be fairly small.
The pressure along these lines of constant physics is compared with
the result from stout-link fermions of Ref.~\cite{AFKS05} in
Fig.~\ref{EOS_2+1f}. With appropriate caveats, the two determinations
agree reasonably well.

A reliable continuum extrapolation of the EOS for full QCD should
become possible in the not too distant future. Certainly, simulations
with $N_t=8$ are needed, and a repeat of the asqtad fermion simulations
for $N_t=4$ and $6$ with the exact RHMC algorithm would be desirable.

\section{Other results}

\begin{figure}[h]
\centering
\includegraphics[width=7cm]{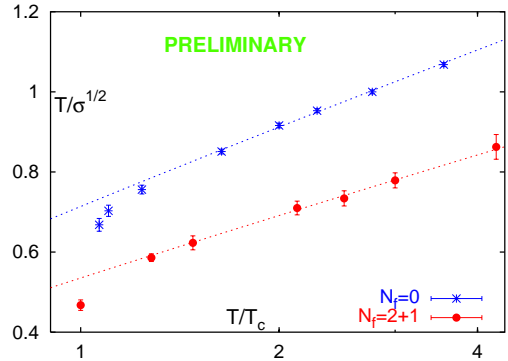}
\caption{\label{sigma_spac} The space-like string tension for pure gauge
theory and for $2+1$ flavor QCD, from Ref.~\cite{Umeda}.}
\end{figure}

At high temperature, one expects dimensional reduction for QCD to work
well, especially in the chromomagnetic sector. The dimensionful 3-d gauge
coupling is given by $g_3^2 = g^2(T) T$, and the fermions, to leading
order, affect only the running of $g(T)$ -- higher order corrections
can be included, see Ref.~\cite{LS05_ss}. This proposition was studied
recently by measuring the spatial string tension in $2+1$ flavor QCD
with p4 fermions~\cite{Umeda} and comparing to pure gauge results,
as shown in Fig.~\ref{sigma_spac}. The comparison can be quantified by
fitting $\sqrt{\sigma} = c_3 g_3^2 = c_3 g^2(T) T$, with $c_3=0.587(41)$
for $2+1$ flavor QCD, as compared to $c_3=0.566(13)$ for 4-d SU(3) pure
gauge theory~\cite{SU3_ss} and $c_3 = 0.553(1)$ for the 3-d
theory~\cite{Teper}. Similarly, for 2-flavor QCD with clover fermions,
Ukita reported $c_3 = 0.54(6)$~\cite{Ukita}, compatible with these results.

Several other interesting results in finite temperature lattice gauge
theory appeared during the last year or were presented at this conference.
These include a study of the dynamics of the phase transition in pure
SU(3) lattice gauge theory~\cite{Bazavov}, indications of a non-trivial
fixed point in finite temperature U(1) lattice gauge theory~\cite{Berg},
a suggestion that the center vortex field is the field that becomes
massless at the deconfinement transition of pure gauge SU(2)
theory~\cite{Langfeld05}, and impressively accurate results for two-color
strong coupling QCD with staggered fermions at finite temperature,
and finite chemical potential, with a novel cluster
algorithm~\cite{Shailesh06}.

\section{Conclusions}

During the last year simulations with improved staggered fermions
have produced convincing evidence that in nature, at zero baryon
density, there is a crossover from the hadronic phase to the quark-gluon
plasma phase, rather than a genuine phase transition~\cite{Szabo}.
A new determination found a crossover temperature of $192(8) \;
\MeV$~\cite{p4_Tc21}, some 10-15\% larger than earlier estimates.
Both results need confirmation from other groups and, in particular,
from other fermion discretizations.

Progress has been made in the computation of the EOS for full QCD at
intermediate temperatures, around the crossover region, where model
and perturbative calculations are unreliable. $N_t=8$ results, though,
are needed for controlled continuum extrapolations.

Much remains to be done in understanding the quark-gluon plasma phase
in the vicinity of the crossover. For example, determinations of transport
coefficients and response functions would be very desirable.

\noindent
{\bf Acknowledgments:} Thanks to all who provided information and figures
for this review, and thanks to my colleagues from the MILC collaboration
and to Frithjof Karsch for numerous discussions on the subject covered here.


\begin{thebibliography}{99}

  \bibitem{Reviews}
  P.~Petreczky,
  Nucl. Phys. Proc. Suppl. {\bf 140}, 78 (2005) [hep-lat/0409139];
  O.~Philipsen,
  Proc. Sci. {\bf LAT2005}, 016 (2005) [hep-lat/0510077].

  \bibitem{Hatsuda}
  T.~Hatsuda,
  these proceedings.

  \bibitem{BHEN}
  A.~Bode, U.M.~Heller, R.G.~Edwards, and R.~Narayanan,
  in ``Structure of the Vacuum'', Kluwer Academic 2000, pg 65
  [hep-lat/9912043];
   R.G.~Edwards, U.M.~Heller and R.~Narayanan,
  Phys. Rev. D {\bf 59}, 094510 (1999) [hep-lat/9811030].

  \bibitem{p4_intro}
  U.M.~Heller, F.~Karsch and B.~Sturm,
  Phys. Rev. D {\bf 60}, 114502 (1999) [hep-lat/9901010].

  \bibitem{ASQTAD}
  Kostas Orginos, Doug Toussaint and R.L.~Sugar,
  Phys. Rev. D {\bf 60}, 054503 (1999) [hep-lat/9903032];
  G.P.~Lepage,
  Phys. Rev. D {\bf 59}, 074502 (1999) [hep-lat/9809157].

  \bibitem{AFKS05}
  Y.~Aoki, Z.~Fodor, S.D.~Katz, and K.K.~Szabo,
  J. High Energy Phys. {\bf 01}, 089 (2006) [hep-lat/0510084].

  \bibitem{MILC04}
  MILC Collaboration: C.~Bernard, {\it et al.},
  Phys. Rev. D {\bf 71}, 034504 (2005) [hep-lat/0405029].

  \bibitem{HMC}
  S.~Duane, A.D.~Kennedy, B.J.~Pendleton, and D.~Roweth,
  Phys. Lett. B {\bf 195}, 216 (1987).

  \bibitem{R_alg}
  S.A.~Gottlieb, W.~Liu, D.~Toussaint, R.L.~Renken, and R.L.~Sugar,
  Phys. Rev. D {\bf 35}, 2531 (1987).

  \bibitem{RHMC}
  M.A.~Clark, B.~Joo and A.D.~Kennedy,
  Nucl. Phys. Proc. Suppl. {\bf 119}, 1015 (2003) [hep-lat/0209035];
  M.A.~Clark and A.D.~Kennedy,
  Nucl. Phys. Proc. Suppl. {\bf 129}, 850 (2004) [hep-lat/0309084];
  M.A.~Clark, A.D.~Kennedy and Z.~Sroczynski,
  Nucl. Phys. Proc. Suppl. {\bf 140}, 835 (2005) [hep-lat/0409133].

  \bibitem{RHMC_lat06}
  M.~Clark,
  these proceedings.

  \bibitem{Shcheredin}
  S.~Shcheredin,
  these proceedings.

  \bibitem{Lombardo}.
  M-P.~Lombardo,
  these proceedings.

  \bibitem{4_th_root}
  S.~Sharpe,
  these proceedings.

  \bibitem{Schmidt}
  C.~Schmidt,
  these proceedings.

  \bibitem{Stephanov}
  M.~Stephanov,
  these proceedings.

  \bibitem{Szabo}
  K.~Szabo,
  these proceedings.

  \bibitem{dFP06}
  Philippe de Forcrand and Owe Philipsen,
  hep-lat/0607017.

  \bibitem{p4_nf3}
  F.~Karsch, E.~Laermann and C.~Schmidt,
  Phys. Lett. B {\bf 520}, 41 (2001) [hep-lat/0107020].

  \bibitem{Cheng}
  M.~Cheng,
  these proceedings.

  \bibitem{KS06}
  J.B.~Kogut and D.K.~Sinclair,
  Phys. Rev. D {\bf 73}, 074512 (2006) [hep-lat/0603021].

  \bibitem{KS00-01}
  J.B.~Kogut and D.K.~Sinclair,
  Phys. Lett. B {\bf 492}, 228 (2000) [hep-lat/0005007];
  Phys. Rev. D {\bf 64}, 034508 (2001) [hep-lat/0104011].

  \bibitem{Pica}
  C.~Pica,
  these proceedings.

  \bibitem{Pisa05}
  M.~D'Elia, A.~Di Giacomo and C.~Pica,
  Phys. Rev. D {\bf 72}, 114510 (2005) [hep-lat/0503030].

  \bibitem{Mendes05}
  T.~Mendes,
  to appear in the Proceedings of the
  ``I Latin American Workshop in High Energy Phenomenology'',
  Porto Alegre, Brazil, December 2005,
  hep-lat/0609035.

  \bibitem{MILC99}
  MILC Collaboration: C.~Bernard, {\it et al.},
  Phys. Rev. D {\bf 61}, 054503 (2000) [hep-lat/9908008].

  \bibitem{p4_Tc21}
  M.~Cheng, {\it et al.},
  hep-lat/0608013, to appear in {\it Phys. Rev.} D;

  \bibitem{Umeda}
  T.~Umeda,
  these proceedings.

  \bibitem{Gray05}
  A.~Gray, {\it et al.},
  Phys. Rev. D {\bf 72}, 094507 (2005) [hep-lat/0507013].

  \bibitem{Tc_coll}
  P.~Petreczky,
  hep-lat/0609040.

  \bibitem{KRT03}
  F.~Karsch, K.~Redlich and A.~Tawfik,
  Eur. Phys. J. C {\bf 29}, 549 (2003) [hep-ph/0303108]

  \bibitem{KLRS03}
  K.~Kajantie, M.~Laine, K.~Rummukainen, and Y.~Schr\"oder,
  Phys. Rev. D {\bf 67}, 105008 (2003) [hep-ph/0211321].

  \bibitem{LS06}
  Mikko Laine and York Schr\"oder,
  Phys. Rev. D {\bf 73}, 085009 (2006) [hep-ph/0603048].

  \bibitem{BDMP06}
  P.~Bialas, L.~Daniel, A.~Morel, and B.~Petersson,
  hep-lat/0606019.

  \bibitem{old_MILC_Nt4}
  T.~Blum, Steven~Gottlieb, Leo~Karkkainen, D.~Toussaint,
  Phys. Rev. D {\bf 51}, 5153 (1995) [hep-lat/9410014].

  \bibitem{old_MILC_Nt6}
  MILC Collaboration: C.W.~Bernard, {\it et al.},
  Phys. Rev. D {\bf 55}, 6861 (1997) [hep-lat/9612025].

  \bibitem{Levkova}
  MILC Collaboration: C.~Bernard, {\it et al.},
  these proceedings.

  \bibitem{Bazavov}
  A.~Bazavov,
  these proceedings.

  \bibitem{Berg}
  B.A.~Berg,
  these proceedings.

  \bibitem{Langfeld05}
  K.~Langfeld, G.~Schulze and H.~Reinhardt,
  Phys. Rev. Lett. {\bf 95}, 221601 (2005) [hep-lat/0508007].

  \bibitem{Shailesh06}
  Shailesh Chandrasekharan and Fu-Jiun Jiang,
  Phys. Rev. D {\bf 74}, 014506 (2006) [hep-lat/0602031].

  \bibitem{LS05_ss}
  M.~Laine and Y.~Schr\"oder,
  Proc. Sci. {\bf LAT2005}, 180 (2005) [hep-lat/0509104].

  \bibitem{SU3_ss}
  G.~Boyd {\it et al.},
  Nucl. Phys. B {\bf 469}, 419 (1996) [hep-lat/9602007].

  \bibitem{Teper}
  B.~Lucini and M.~Teper,
  Phys. Rev. D {\bf 66}, 097502 (2002) [hep-lat/0206027].

  \bibitem{Ukita}
  N.~Ukita {\it et al.},
  these proceedings.

\end{thebibliography}
\end{document}